\documentclass[sigconf]{acmart}

\AtBeginDocument{%
  \providecommand\BibTeX{{%
    \normalfont B\kern-0.5em{\scshape i\kern-0.25em b}\kern-0.8em\TeX}}}

\copyrightyear{2021}
\acmYear{2021}
\setcopyright{acmlicensed}\acmConference[KDD '21]{Proceedings of the 27th ACM SIGKDD Conference on Knowledge Discovery and Data Mining}{August 14--18, 2021}{Virtual Event, Singapore}
\acmBooktitle{Proceedings of the 27th ACM SIGKDD Conference on Knowledge Discovery and Data Mining (KDD '21), August 14--18, 2021, Virtual Event, Singapore}
\acmPrice{15.00}
\acmDOI{10.1145/3447548.3467087}
\acmISBN{978-1-4503-8332-5/21/08}

\usepackage[colorinlistoftodos]{todonotes}
\usepackage{parskip}
\usepackage{multirow}
\usepackage{xspace}
\usepackage{subcaption}
\usepackage{booktabs}
\usepackage{dblfloatfix} 
\usepackage{soul}


\newcommand{\xmr}{\textsf{XMR}\xspace}
\newcommand{\mfq}{\textsf{MFQ}\xspace}
\newcommand{\mfqplus}{\textsf{MFQ + seq2seq-GRU}\xspace}
\newcommand{\qac}{\textsf{QAC}\xspace}

\newcommand{\parabel}{\textsf{Parabel}\xspace}
\newcommand{\pecos}{\textsf{PECOS}\xspace}
\newcommand{\our}{\textsc{PrefXMRtree}\xspace}
\newcommand{\GRU}{\textsf{GRU}\xspace}
\newcommand{\RR}{\mathbb{R}}
\newcommand{\unigram}[1]{\phi_{\text{tfidf}}\left(#1\right)}
\newcommand{\charngram}[1]{\phi_{\text{c-tfidf}}\left(#1\right)}
\newcommand{\labelText}{\phi_{\text{c-tfidf}}(\ell_{\text{text}})}
\newcommand{\pifa}{\phi_{\text{pifa}}(\ell)}
\newcommand{\concat}{\mathbin\Vert}
\newcommand{\prevQuery}{q_l}
\newcommand{\prefix}{p}
\newcommand{\RRBLEU}{\textsc{Bleu\textsubscript{rr}}\xspace}
\newcommand{\BLEU}{\textsc{Bleu}\xspace}
\newcommand{\nextQuery}{q^\star}
\newcommand{\prevQueryEmbed}{\phi_{\text{tfidf}}\left(\prevQuery\right)}
\newcommand{\prevQueryPrefixEmbed}{\phi_{\text{tfidf}}\left(\prevQuery\right) \concat \phi_{\text{c-tfidf}}\left(\prefix \right)}
\newcommand{\aol}{AOL search logs\xspace}

\begin{document}

\title{Session-Aware Query Auto-completion using Extreme Multi-Label Ranking}


\author{Nishant Yadav}
\authornote{Work done while at Amazon}
\affiliation{
  \institution{University of Massachusetts Amherst}
  % \country{USA}
}

\author{Rajat Sen}
\authornotemark[1]
\affiliation{
  \institution{Google Research}
  % \country{USA}
}

\author{Daniel N. Hill}
\affiliation{
  \institution{Amazon}
  % \country{USA}
}

\author{Arya Mazumdar}
\authornotemark[1]
\affiliation{
  \institution{University of California, San Diego}
  % \country{USA}
}

\author{Inderjit S. Dhillon}
\authornotemark[1]
\affiliation{
  \institution{ Amazon \& University of Texas Austin}
  % \country{USA}
}

\renewcommand{\shortauthors}{Yadav, et al.}

\begin{abstract}
Query auto-completion (QAC) is a fundamental feature in search engines 
where the task is to suggest plausible completions of a prefix typed 
in the search bar. Previous queries in the user session can provide 
useful context for the user’s intent and can be leveraged to suggest 
auto-completions that are more relevant while adhering to the user’s 
prefix. Such session-aware QACs can be generated by 
recent sequence-to-sequence deep learning models; however, these 
generative approaches often do not meet the stringent latency 
requirements of responding to each user keystroke. Moreover, these 
generative approaches pose the risk of showing nonsensical queries. 
One can pre-compute a relatively small subset of relevant queries for 
common prefixes and rank them based on the context. However, such an 
approach fails when no relevant queries for the current context are 
present in the pre-computed set. 

In this paper, we provide a solution to this problem: we take the novel 
approach of modeling session-aware QAC as an 
eXtreme Multi-Label Ranking (XMR) problem where the input is the previous 
query in the session and the user’s current prefix, while the output space 
is the set of tens of millions of queries entered by users in the recent past. 
We adapt a popular XMR algorithm for this purpose by proposing several 
modifications to the key steps in the algorithm. The proposed modifications 
yield a 3.9$\times$ improvement in terms of Mean Reciprocal Rank (MRR) over the baseline 
XMR approach on a public search logs dataset. We are able to maintain an inference
latency of less than 10 ms while still using session context. When compared against 
baseline models of acceptable latency, we observed a 33\% improvement in MRR for short prefixes
of up to 3 characters. Moreover, our model yielded a statistically significant 
improvement of 2.81\% over a  production QAC system in terms of suggestion acceptance rate, when 
deployed on the search bar of an online shopping store as part of an A/B test. 
\end{abstract}


\begin{CCSXML}
<ccs2012>
   <concept>
       <concept_id>10002951.10003260.10003261.10003271</concept_id>
       <concept_desc>Information systems~Personalization</concept_desc>
       <concept_significance>500</concept_significance>
       </concept>
   <concept>
       <concept_id>10010147.10010257.10010258.10010259.10003268</concept_id>
       <concept_desc>Computing methodologies~Ranking</concept_desc>
       <concept_significance>500</concept_significance>
       </concept>
 </ccs2012>
\end{CCSXML}

\ccsdesc[500]{Information systems~Personalization}
\ccsdesc[500]{Computing methodologies~Ranking}

\keywords{eXtreme multi-label ranking, auto-complete, session-aware}


\maketitle

\section{Introduction}
Query auto-completion (\qac) is an important component of modern search engines. 
The task is to generate query suggestions that start with the
prefix typed by the user in the search bar. This is of great utility to users as 
it saves them the effort of typing out the 
complete query. \qac also helps the user to formulate a query when they are not 
completely sure what to search for. Often a user will make a series of queries within a small time window,
referred to as a \textit{session}. 
In such cases, the user's previous 
queries provide contextual information that can be used to 
make their auto-complete 
suggestions more relevant~\cite{bar2011context, sordoni2015hierarchical, jiang2014learning}. Consider an example 
where the user has typed \texttt{"n"} into the search bar after previously searching for \texttt{"digital camera"}. 
Based on the previous query, it is more likely 
that the user is searching for \texttt{"nikon camera"} than for \texttt{"nike shoes"}. Thus the suggestion
\texttt{"nikon camera"} should be ranked higher than \texttt{"nike shoes"} in the list of suggestions. 
In this paper, we propose a novel method for session-aware \qac which uses contextual information from previous queries made by 
the user to suggest better query completions within a strict latency budget.

\qac is typically performed in a two-step retrieve-and-rank 
process. In the first step, a limited number of candidate queries 
that begin with the user's prefix are retrieved. These candidates are pre-selected 
from prior data based on business rules. For instance, one could select the top 
100 most frequent completions for a prefix based on prior data. 
These retrieved queries are then ranked based on 
features like frequency, similarity to the previous query, 
user profile, etc.~\cite{sordoni2015hierarchical,cai2014time,
shokouhi2013learning,wang2020efficient}. Finally, 
the top-$k$ ranked queries are shown as suggestions to the user. 
However, this approach
fails when the candidate set does not contain relevant queries for the given context. 
This problem is especially severe for very short prefixes where  
the limited and fixed candidate set is unlikely
to contain relevant queries for all possible contexts.

To alleviate the shortcomings of the retrieve-and-rank 
framework, another line of research uses generative sequence 
models to \emph{generate} potential query completions, starting 
from the given prefix and conditioned on previous 
queries~\cite{wang2018realtime,dehghani2017learning, song2017hierarchical,jiang2018neural}. 
Such models leverage powerful 
deep learning models that can generate novel queries as well as generalize to 
unseen prefixes. However, these generative sequence models typically have high 
inference latencies that prohibit their use in real-time systems where the 
allowed latency budget is on the order of few milliseconds to match 
the user's typing pace~\cite{wang2018realtime}. Moreover, using a generative 
model increases the risk of surfacing non-sensical queries in business-critical applications. 

To overcome the limitations of existing techniques while still meeting the low 
latency requirements, we take the novel approach of modeling 
session-aware query auto-completion using an 
\textit{end-to-end} retrieve-and-rank framework of eXtreme Multi-label Ranking (\xmr) 
models. \xmr algorithms are designed to match and rank a list of relevant 
items from a huge, fixed
output space containing millions of labels. This approach ranks
a \emph{context-dependent} list of candidate suggestions matched from the 
huge output space, in contrast to retrieve-and-rank methods with a fixed 
context-independent list of candidates. Also, note that the
output space of suggestions in \xmr can be pre-filtered to reduce the risk 
of surfacing non-sensical queries, unlike in generative approaches.

Session-aware \qac can be framed as a multi-label ranking task where the 
input is the user's prefix and previous queries, and the  
observed next query is the ground-truth label. Several promising 
approaches for \xmr are based on the concept of forming a probabilistic
label tree~\cite{jasinska2016extreme, prabhu2018parabel, khandagale2020bonsai, you2019attentionxml, pecos2020}. 
We show that such tree-based methods are especially amenable to be adapted for the \qac problem.
In this work, we adapt \parabel~\cite{prabhu2018parabel}, a popular 
tree based \xmr model, for this task. In particular, we use a flexible and 
modular open-source \xmr framework, \pecos~\cite{pecos2020}, which generalizes the \parabel model. 
\pecos proceeds in three steps: (i) \textit{Label Indexing.} Organize the large 
number of labels in a hierarchical label tree. (ii) \textit{Matching. } Given an 
input context, retrieve a much smaller subset of labels of interest. 
(iii) \textit{Ranking.} Rank the relevant labels to return the top-$k$ items. 
\pecos leverages the hierarchical structure of label indexing to perform 
matching and ranking with inference time logarithmic in the total number of 
labels. A straightforward way to apply  \pecos to our task is to use the previous 
query to generate a ranked list of next query suggestions and filter out any suggestion that 
does not match the prefix. However, using \pecos out-of-the-box performs poorly. 
In this work, we adapt the indexing, matching, and ranking  steps in \pecos so that 
the model retrieves and ranks queries (labels)
matching the input prefix with higher probability without compromising on suggestion 
quality and inference latency.
In summary, the key contributions of this paper are:
\begin{itemize}
	\item To the best of our knowledge, this is the first work to use 
	an end-to-end retrieve-and-rank \xmr framework for the session-aware 
	query auto-complete problem. 
	\item We propose a set of label indexing methods for tree-based \xmr 
	which are amenable to the \qac problem. In particular, one of 
	our label indexing algorithms uses a hybrid of a trie 
	with 2-means hierarchical clustering based on a novel 
	position-weighted TF-IDF vectorization of the label text. 
	This leads to an improvement of 26\% in MRR on the \aol 
	dataset~\cite{pass2006picture}, over the default indexing scheme in \pecos.
	\item  On the \aol dataset, the proposed \xmr model yields up to 
	3.9$\times$ improvement over using \pecos out-of-the-box. It outperforms all baseline approaches 
	that meet latency requirements (less than 10 ms per inference),
	and outperforms all baselines when the ground-truth label is present in the 
	output space. Our gains over baseline models are sharpest for shorter prefixes 
	where the context information is extremely important.  
	For prefixes of 3 or fewer characters, we see a 
	33\% improvement over the best baseline that meets latency requirements.
	\item Our model yielded a statistically significant improvement of 2.81\% over a production 
	system in terms of QAC suggestion acceptance rate when 
	deployed on the search bar of an online shopping store as part of an A/B test. 
\end{itemize}

\section{Related Work}

Traditional query auto-completion systems operate in the retrieve-and-rank framework where 
a fixed set of queries are retrieved given the input prefix
followed by ranking based on various sources of information. 
The simplest yet effective approach is to suggest the top-$k$ most frequent queries matching 
the input prefix~\cite{wang2020efficient}. Other approaches further improve over this by 
ranking suggestions based on previous queries, user profile, time-sensitivity, or coherence 
with the typed prefix~\cite{sordoni2015hierarchical,cai2014time,
shokouhi2013learning,wang2020efficient}.
Recent work also leverages deep learning models to generate additional features such as 
query likelihood using language models~\cite{park2017neural},  
personalized language models~\cite{fiorini2018personalized, jaech2018personalized}, 
or previous queries~\cite{sordoni2015hierarchical} to rank queries using 
a learning to rank framework~\cite{wu2010adapting}.
However, these models typically rely on \emph{heuristic} methods to generate a fixed 
number of candidates for all popular prefixes. Our work differs from these in that we propose an 
end-to-end model comprising of a \emph{learned} candidate query retrieval which is trained 
\emph{jointly} with a query ranker. Moreover, our candidate set is dynamic and 
is fetched from an enormous output space as a function of the context.

Another line of research uses seq2seq and language models to \emph{generate} query 
suggestions given a prefix, and can generalize to unseen prefixes and potentially 
generate new queries.
\citet{wang2018realtime} use character-level language models to generate completions for 
a given prefix, additionally combining it with a spelling correction module.
\citet{dehghani2017learning} augment a seq2seq model with attention mechanism to attend to 
promising parts of previous queries during decoding together with copy mechanism to copy 
words from previous queries. ~\citet{mustarusing} fine-tunes pre-trained language models 
such as BERT~\cite{devlin2018bert} to generate auto-complete suggestions.
However, unlike models proposed in this work, these models typically have high inference latency which 
prohibits their use in realtime systems, and could potentially generate non-sensical auto-complete suggestions.

In general, \xmr models fall into three categories: 
1-vs-all approaches \cite{yen2016pd, yen2017ppdsparse, weston2013label, babbar2017dismec}, 
embedding based methods~\cite{malkov2018efficient, bhatia2015sparse, chen2012feature, cisse2013robust, mineiro2015fast, lin2014multi}, 
and tree-based methods~\cite{jasinska2016extreme, prabhu2018parabel, khandagale2020bonsai, you2019attentionxml, pecos2020}. 
A recent work~\cite{jain2019slice} formulates next query suggestion 
\emph{without prefix constraints} as an \xmr task, and proposes Slice, an \xmr algorithm 
based on navigable small world graphs~\cite{malkov2018efficient}. Slice improves on 
limitations of traditional related search suggestion approaches in terms of coverage and 
suggestion density but it is non-trivial to extend Slice to adhere to prefix constraints 
in a time-efficient manner for query auto-completion.
Tree-based data structures such as tries and ternary trees have been widely used in query auto-completion 
systems~\cite{xiao2013efficient, hsu2013space, kastrinakis2010advancing, cai2016query}. 
Trees offer an inductive bias for handling the variable length prefix constraint in a time and 
space efficient manner. For this reason, we build on top
of tree-based \xmr methods in this work.

\section{Problem Formulation}
Users often submit a sequence of queries to the search 
engine in a single session. These recent queries provide useful contextual 
information to improve the auto-complete suggestions shown to the user. 
The task is as follows: 
\textit{given the last 
query $\prevQuery$ made by the user, and the next query's prefix $\prefix$ typed by 
the user so far, generate a ranked list of top-$k$ next query suggestions 
that match the prefix given by the user and maximize $\hat{\mathbb{P}}$(next query| previous query, prefix).}
In this work we use only the most recent query in the session for 
generating relevant auto-complete suggestions. However, the techniques 
introduced can be easily extended to use more than one previous query from 
the user as context. 

In this work, we cast session-aware query auto-completion as an extreme 
multi-label ranking (\xmr) problem and use tree-based algorithms developed 
for \xmr which allow time-efficient inference, while handling a large 
output space.

Given $N$ training points, 
$\{x_i, y_i\}_{i=1}^{N}, x_i \in \RR^{d}, y_i = \{0,1\}^{L}$, 
where $L = |\mathcal{L}|$ and $\mathcal{L}$ is the set of labels  
(next queries in our case), an \xmr model 
learns a function $f : \RR^{d} \rightarrow [0,1]^{L}$. While \xmr models might be able to 
generate a relevance score for every label given the input, only a small subset 
of labels are active for a data point. Therefore, \xmr models typically 
generate a ranked list of top-$k$ labels, $k \ll L$, given an input. 
We will now describe the specific \xmr algorithm that we build upon.
 
\subsection{The \pecos Framework}
In this work, we adapt the flexible and modular \xmr framework 
called \pecos~\cite{pecos2020} for the \qac problem. One instance of 
the \pecos framework known as XLinear generalizes a popular tree-based \xmr 
algorithm called \parabel~\cite{prabhu2018parabel}. We use this instance 
of \pecos as a starting point. In what follows \pecos will mean the XLinear 
version of \pecos, which is conceptually equivalent to \parabel. 
At a high level, \pecos has three components:
\begin{itemize}
	\item A {\bf label indexing} model which organizes labels in 
	hierarchical clusters by grouping similar
	labels together.
	\item A {\bf matching} model which retrieves a small subset of label 
	clusters given an input.
	\item A {\bf ranking} model which ranks labels in the  subset 
	of label clusters retrieved by the matching model for the given input.
\end{itemize}

\pecos organizes labels in a label tree with all the labels present in leaf nodes, 
and learns a probabilistic model of the joint label distribution. The label
tree is created by recursively partitioning labels into $k_b$ balanced groups
until each group/cluster contains less than $M$ labels. 
\pecos learns a 1-vs-all classifier at each internal node of the label tree
as well as a 1-vs-all classifier for each label in each leaf node.
At inference time, a given input $x$ is routed to leaf nodes
using the 1-vs-all classifiers at internal nodes. It starts from the root node and uses
1-vs-all classifiers at internal nodes to decide whether the data point should
proceed further along a branch. Therefore, the data point can traverse multiple paths
and reach multiple leaf nodes. In order to avoid reaching all leaf nodes in the worst case, \pecos
greedily selects up to $B$ leaf nodes, when using beam search with beam width $B$.
Beam search starts with the root node and at each level, it evaluates the probability of reaching child nodes of
each node in the beam, selects up to $B$ most probable child nodes, and proceeds to the next level.
Beam search terminates with up to $B$ most probable leaf nodes in 
time $O(B\hat{D}k_b\log L$) where $\hat{D}$ is average label feature 
sparsity ($\hat{D} = D$ for $D$-dim dense features).
Finally, the relevance score of each label $\ell$ in the retrieved leaf nodes is 
calculated using its 1-vs-all classifier together with the probability of 
path leading to its leaf node. Since there are at most $M$ labels in each 
leaf node, this step takes $O(B\hat{D}M)$ time, and the overall inference time 
complexity is $O(B\hat{D}( k_{b}\log L + M))$. 

\paragraph{PECOS out-of-the-box for \qac}
\label{sec:indexing}

\begin{figure*}[ht!]
  \centering
  \includegraphics[trim=0 0 0 0, clip,width=\linewidth]{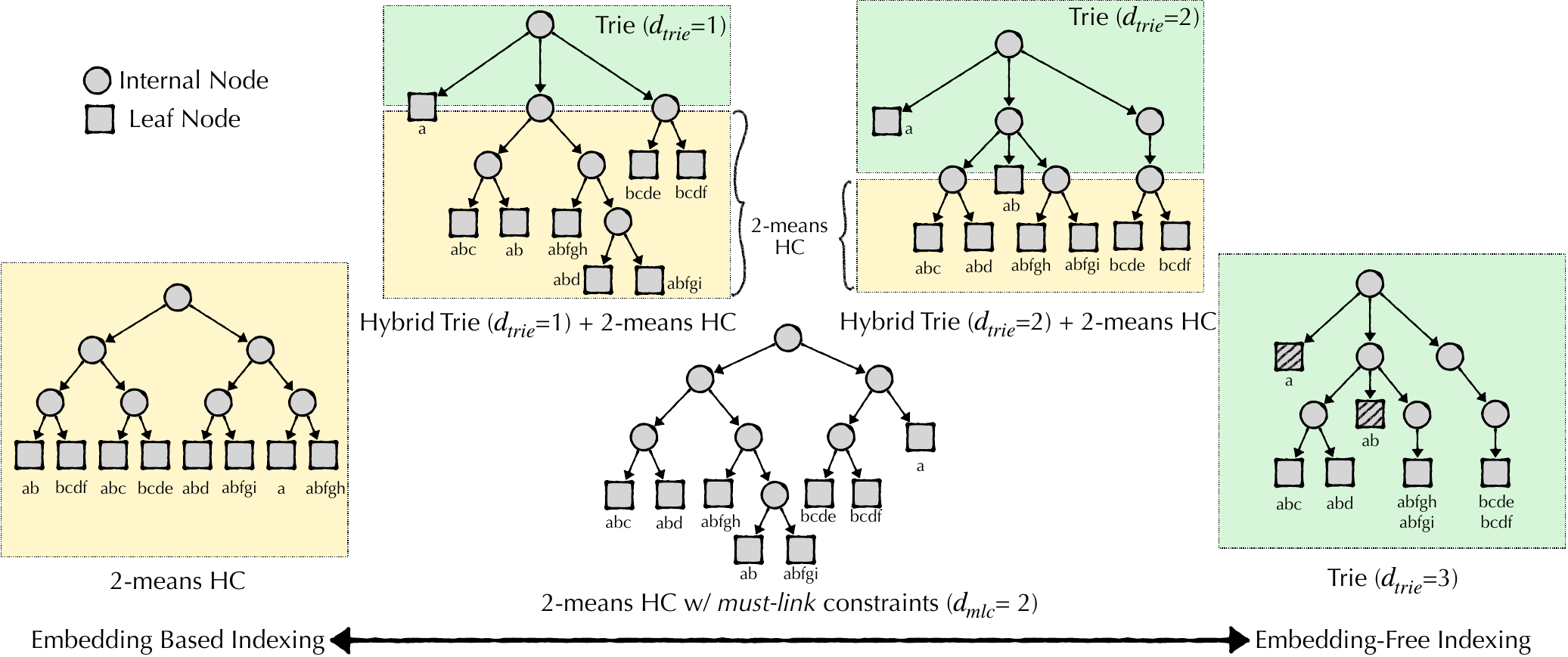}
  \caption{Label indices on a set of eight labels. We have embedding-based 
  indexing methods such as 2-means hierarchical clustering (HC) on the left, and
  embedding-free indexing using a trie on the right. In the middle, we 
  have indices which interpolate between the two extremes such as hybrid trie indices and 
  index build using 2-means HC with \emph{must-link} constraints.}
  \label{fig:label_indices}
\end{figure*}

A simple way to use \pecos for this task is a two step
approach of ranking followed by filtering. First, we can use \pecos 
to predict a ranked list of labels (next queries) given the previous query $\prevQuery$ 
as input, and use the prefix $\prefix$ to collect suggested next queries that 
match the prefix. Since the number of labels is typically of the order of millions,
\pecos ranks a small subset of labels for efficiency. This 
small subset of labels often contains very few labels (next queries) 
matching the prefix when using \pecos out-of-the-box, therefore often 
resulting in an empty list of auto-complete suggestions. One simple solution 
to counter this issue is to rank a much \emph{larger} fraction of label 
space by increasing 
beam width ($B$) used in beam search at test time. 
However, increasing $B$ causes a corresponding increase 
in the inference latency, potentially rendering it unsuitable 
for deploying in latency-critical systems.

To alleviate this issue, we make each component of \pecos 
\textit{prefix-aware} so that the model retrieves and ranks next queries 
(labels) matching the input prefix with higher probability without
compromising on inference latency and suggestion quality. We use \our to refer to 
\pecos \xmr models with label indexing, matching and ranking models proposed in this work.

\subsection{Label Indexing in \our}
The label indexing step in \xmr tree models aims to organize labels into hierarchical 
clusters by grouping similar
labels together. For example, \parabel indexes labels using spherical 
2-means hierarchical clustering with label 
embeddings obtained using positive instance feature aggregation (PIFA). 
The matching model navigates through the index and 
retrieves a small subset of label clusters. Then, the ranking model 
ranks the labels present in those clusters.
For this reason, the label index is \our's backbone in that the 
performance of the matching model, and in turn, the ranking model relies on 
the quality of the index to a large extent. 

\subsubsection{Label Indexing Algorithms}
We now propose different ways of constructing   
tree-based indexes over the label space using approaches ranging from 
embedding-based methods such as hierarchical 2-means, to embedding-free 
approaches such as a \emph{trie}, and hybrid approaches that interpolate between 
the two extremes. We also explore different label embedding strategies.

\paragraph{Spherical 2-means Hierarchical Clustering}
This is the default indexing algorithm in \pecos. The
labels are first embedded in $\RR^p$, unit normed and then clustered 
with 2-means hierarchical clustering using cosine similarity subject to 
balance constraints. The balance constraint ensures that for a given 
parent cluster, the size of two child clusters output by 2-means are 
at most one unit apart in size. The recursive partitioning of an internal 
node is stopped when the node has less than $M$ labels, and thus  
resulting in a cluster tree of depth $\mathcal{O}(\log (L/M))$.

\paragraph{Trie}
The labels (next queries) are organized in a trie data 
structure using the label string. The depth of a trie is 
proportional to length of the longest string stored in it which can be 
much larger than average length of its data strings. So, we limit the depth 
of the final trie index to a given depth $d_{trie}$ and collapse all subtrees at 
depth $d_{trie}$ to a single leaf node containing all the labels in the 
corresponding collapsed subtree. \pecos assumes that all the labels
are stored in leaf nodes of the index tree but some labels might be stored 
at internal nodes of a trie. For instance, if we have two labels 
\texttt{"ab"} and \texttt{"abc"}
and the depth of trie index is limited to 3, then \texttt{"ab"} would be stored 
at an internal node on path to leaf containing \texttt{"abc"}. 
In such cases, we create an additional dummy leaf node as a child 
to the internal node storing the label, \texttt{"ab"} in this example, 
and move the label to the dummy leaf node.  
In Figure ~\ref{fig:label_indices} on the far right, 
we show dummy leaf nodes created for labels \texttt{"a"} and \texttt{"ab"}.

While trie index perfectly groups labels by prefixes, the resulting 
hierarchical index structure has a \emph{high} branching factor, as well as 
a \emph{skewed} distribution of the number of labels in each leaf node. Our
initial experiments revealed that this 
adversely affects the training time as well as inference latency of 
\our with a trie index but it outperforms hierarchical clustering 
based indices in terms of output quality. The primary reason behind 
trie's superior statistical performance for this task is its ability to group 
labels with common prefixes together. On the other hand, hierarchical 
clustering based indices have better training time and inference latency 
because of their balanced structure. In order to achieve the
best of both worlds, we propose the following two indexing strategies to
interpolate between the two extremes.

\paragraph{Hierarchical Clustering with \emph{must-link} constraints}
The labels are organized using spherical 2-means hierarchical clustering (as discussed before) 
with additional \emph{must-link} constraints at each level
to ensure that labels with the same prefix up to length 
$d'$ are present in the same cluster at depth $d'$ of the index. The goal of these
constraints is to mimic trie like behaviour of grouping labels with common prefixes,
and the use of 2-means hierarchical clustering yields a balanced index with a limited branching 
factor of 2. Imposing \emph{must-link} constraints for all levels of the 
index suffers from trie like imbalance in depth of the index. Therefore, 
these must-link constraints are enforced up to a fixed depth 
$d_{mlc}$. 
Figure~\ref{fig:label_indices} shows in the lower center an example of clustering with 2-means hierarchical
clustering with must-link constraints up to depth $d_{mlc}=2$. Note that in an attempt
to build a balanced tree with a fixed branching factor, it merges branches containing 
labels \texttt{"a"}, and {\texttt{"bcde"} and \texttt{"bcdf"}}.
Labels with same prefix up to length 2  such as \texttt{"abfgh"} and \texttt{"abfgi"} 
are constrained to be in the same subtree up to depth 2 but can be present
in different subtrees at depth 3 as must-link constraints are only enforced up to depth 2 in this example.

\paragraph{Hybrid Indexing}
This approach interpolates between 2-means hierarchical clustering and 
trie based indexing by organizing labels in a trie of depth $d_{trie}$, and then 
further clustering labels present in each branch using 2-means 
hierarchical clustering. The resulting index has a relatively high 
branching factor and imbalanced label load distribution up to depth $d_{trie}$ 
and a balanced index structure with a branching factor of 2 beyond that.
Figure~\ref{fig:label_indices} shows a hybrid trie index in which labels
are organized using a trie for a fixed depth $d_{trie}$= 1 and $d_{trie}$= 2, followed by 2-means
hierarchical clustering in each subtree from depth $d_{trie}$ onward.

\subsubsection{Label Embedding}
\label{subsubsec:label_embedding}
Indexing algorithms such as $k$-means hierarchical clustering require label 
embeddings in order to index them. We generate label embeddings using the
following strategies:
\begin{itemize}
	\item Positive Instance Feature Aggregation (PIFA) Embeddings: Given $n$ 
	training points $\{x_i, y_i\}_{i=1}^{n}$, where $x_i \in \RR^d$, and 
	$y_i \in \{0,1\}^{L}$, a label $\ell$ is embedded as:
	\[ \pifa = v_\ell/\| v_\ell \|, \text{where } v_\ell = \sum_{i=1}^{N} y_{i\ell} x_i \]
	Thus, the embedding of a label is the unit-normed average embedding 
	of training data points for which the given label is active. 
	Different types of input embeddings (discussed in 
	section~\ref{subsec:adapt_matching_ranking}) yield 
	different label embeddings using PIFA. 
	\item Label Text Embedding: Given a label $\ell$, we embed the 
	label using \textit{character} n-gram tfidf vector of the label text. 
	We use simple tfidf vectors~\cite{jones1972statistical} as well as novel position-weighted 
	tfidf vectors for generating label embeddings from label text.
\end{itemize}

Simple tfidf vectorization of a label (next query) ignores the position of  
character n-grams in it. For example,
\texttt{"nike shoes"} and \texttt{"shorts nike"} contain similar character 
n-grams but occurring at different positions.
Label indexing methods using such embeddings often fail to 
group labels with similar prefixes together.  As a result, matching and ranking 
models might retrieve and rank labels that do not match the prefix (but might have 
similar character n-grams). This hurts overall 
performance as most of the labels get eventually filtered out because of a prefix 
mismatch. Instead, we want to change the label embedding such that 
labels with similar prefixes are grouped together during the indexing step so that matching 
and ranking models are more likely to retrieve and give higher rank to labels 
matching the given prefix.

\begin{figure}[ht!]
  \centering
  \includegraphics[trim=0 0 0 0, clip,width=\linewidth]{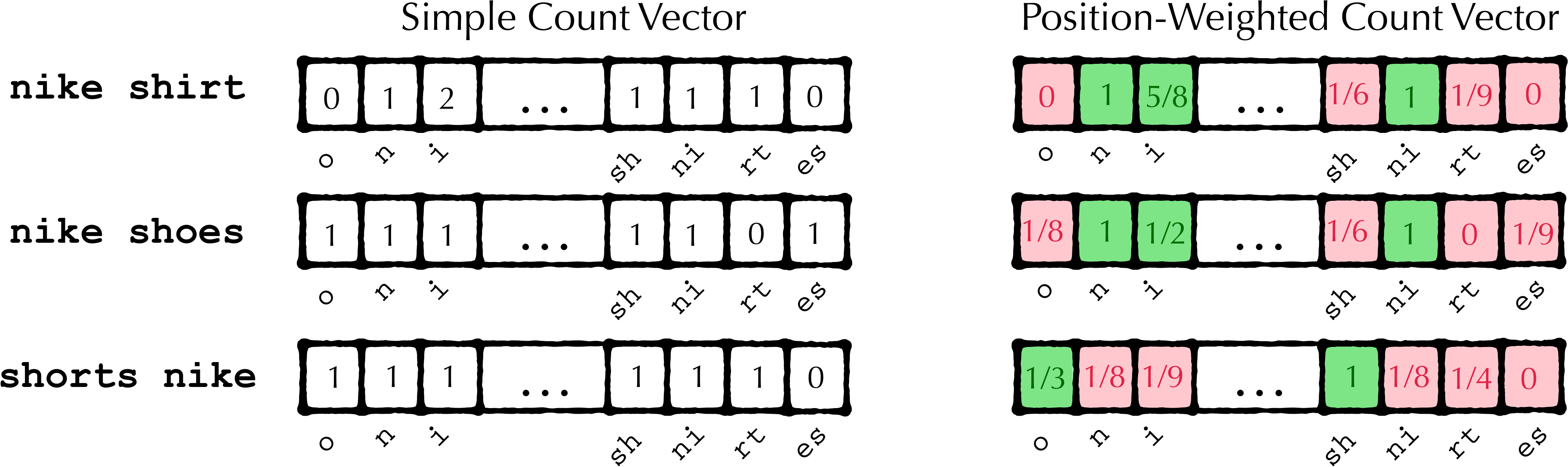}
  \caption{This example illustrates that position-weighting leads to \texttt{"nike shorts"} 
  and \texttt{"nike shoes"} being much closer as vectors than \texttt{"shorts nike"}.}
  \label{fig:pos_wgtd_eg}
\end{figure}

\paragraph{Position-Weighted tfidf vectors}
Tfidf vectorization first counts occurences of an n-gram in the input 
and then multiplies it with its inverse document frequency to get the final 
score for each n-gram. Intuitively, we wish to assign a higher weight 
to n-grams which occur towards the beginning of the prefix than those 
which occur towards the end so that labels with similar prefixes have 
higher similarity than those having merely common character n-grams 
(possibly at different positions in the labels). 
We make tfidf vectorization position sensitive by modifying its counting step. 
Instead of each occurrence of n-gram contributing 1 unit towards its count, 
we modify it to contribute inversely proportional to its position. 
Specifically, the occurrence of an n-gram at position $i$ contributes 
$1/{i}$ towards its count. Figure~\ref{fig:pos_wgtd_eg} shows character
n-gram count vectors created during simple and position-weighted tfidf vectorization 
for \texttt{"nike shoes"}, \texttt{"nike shirt"}, and \texttt{"shorts nike"}. Simple count vectors
look similar for all three examples but position-weighted count vectors
for \texttt{"nike shoes"} and \texttt{"nike shirt"} are more similar than 
those for \texttt{"nike shoes"} and \texttt{"shorts nike"}.
The final count of each n-gram thus obtained is multiplied with its inverse 
document frequency as in simple tfidf vectorization.

\subsection{Encoding the Input for Matching and Ranking in \our}
\label{subsec:adapt_matching_ranking}
The input embedding is used by the matching model to retrieve relevant label clusters
and the ranking model to rank the labels in the retrieved clusters. 
The input consists of the previous 
query ($\prevQuery$), and prefix ($\prefix$) typed by the user. A 
straightforward way of featurizing the input is to only use the
token-level tfidf vector of the previous query ($\prevQuery$) and 
ignore the prefix $(\prefix)$.
\[ x = \phi_{\text{tfidf}}\left(\prevQuery\right)  \] 
However, this featurization may lead to retrieval of very few 
labels which match the given prefix. In order to make 
the matching and ranking model retrieve and rank labels 
starting with the given prefix with higher probability, we generate input embedding by concatenating 
a token based unigram tfidf embedding of $\prevQuery$ with a character n-gram
tfidf embedding of $\prefix$. 
\[ x = \phi_{\text{tfidf}}\left(\prevQuery\right) \concat \phi_{\text{c-tfidf}}\left(\prefix \right) \] 
For embedding 
the prefix, we experiment with simple tfidf vectors as well as position-weighted 
tfidf vectors proposed in Section~\ref{subsubsec:label_embedding}.

\section{Experimental Results}

We evaluate several \our models built with different combinations of label 
indexing methods and input embeddings on \aol 
dataset~\cite{pass2006picture} and a dataset from an online shopping store, and compare against various popular baselines. 
We first study the performance of an out-of-the-box \pecos model and our proposed \our variants, and then 
compare the best performing \our models with generative seq2seq models, and 
query-frequency based auto-complete baselines. 

\paragraph{Dataset}

\aol consist of sequences of queries made by the user. 
we lowercase all queries, replace periods with spaces, and remove non-alphanumeric 
characters. Further, we split a sequence of queries into sessions using a 30-min idle time 
between consecutive queries to define the end of a session~\cite{sordoni2015hierarchical,wang2020efficient}.
Given a session consisting of a sequence of queries 
$<q_1, \ldots, q_{k}>$, we generate $k-1$ (previous query, prefix, next query) 
triplets. The $i^\text{th}$ triplet is 
$(q_i, p_{i+1}, q_{i+1})$, where $p_{i+1}$ is a uniformly sampled prefix of $q_{i+1}$.
Each such triplet serves as a data point where the input is the previous 
query and prefix pair and the ground-truth output is the next query.
Table~\ref{tab:aol_stats} shows statistics of train/test/dev splits for \aol dataset.

We also run experiments using search logs from an online shopping store. Like \aol, 
we split the data into train/test/dev based on timestamp, using the first few weeks' data 
for training, and the next few weeks' data for dev and test. The training data consist of
180MM data points with 23MM unique next queries. The test and dev
set consists of 100K data points each, with 11K unique next queries of which 60\% next queries
were seen at training time.

\paragraph{Evaluation}
We evaluate all proposed models using mean reciprocal rank (MRR) and  
reciprocal rank weighted average of the \BLEU~\cite{papineni2002bleu} score (\RRBLEU)
of the predicted top-$k$ auto-complete 
suggestions.
In our experiments, we use $k=10$.

We also measure the inference latency of each model and report 50th percentile ($p50$) and 99th percentile ($p99$) of inference latencies over the test set. All our \our models are trained and evaluated using a machine with Intel(R) Xeon(R) Platinum 8259CL @ 2.50GHz CPU with 16 cores. The seq2seq-GRU model is trained and evaluated on the more expensive NVIDIA V100 Tensor Core GPU with 32 GB memory.

\subsection{Baselines}
\begin{itemize}
	\item seq2seq-GRU : An RNN model with GRU~\cite{cho2014learning} encoder model is trained 
	to encode the previous query $\prevQuery$, and the encoded representation 
	is fed in to an RNN decoder with GRU units to generate the next query.
	\item Most-Frequent Query (\mfq): This model returns top-$k$
	most frequent suggestions matching the input prefix $\prefix$.
	This is a strong baseline
	corresponding to the maximum likelihood estimator for $P$(next query | prefix)~\cite{wang2018realtime}. 
	Note that since this model completely ignores the previous query $\prevQuery$, 
	list of suggestions for each prefix can be pre-computed and stored in a database.
	\item \mfqplus: This model retrieves $10k$ suggestions using \mfq and uses a seq2seq-GRU model
	to rank suggestions based on the conditional probability of each suggestion given 
	previous query $\prevQuery$, and returns top-$k$ suggestions.
\end{itemize}
\begin{table}[!t]
  \caption{Statistics of AOL Search Logs Dataset}
  \label{tab:aol_stats}
  \begin{tabular}{lccc}
    \toprule
                                      & Train    & Dev    & Test\\
    \midrule
     Date Range (2006)                & 3/1 - 5/15     &  5/16 - 5/23       & 5/24 - 5/31 \\
     \# Sessions                      & 3,569,171         &  325,262          & 316,640               \\ 
     \# Query Pairs                   & 9,275,412         &  830,227          & 803,366               \\ 
     \# Unique Next Queries           & 5,581,896         &  615,496          & 590,486             \\ 
  \bottomrule
\end{tabular}
\end{table}

We refer the reader to Appendix~\ref{sec:implementation_details} for more details on 
the evaluation metrics, data preprocessing, and model implementation. 
\footnote{Code for reproducing experiments
is available at \url{https://github.com/amzn/pecos}.}

 \begin{table*}[!ht]
  \caption{Mean Reciprocal Rank (MRR), inference latency and training time of \our models for different combinations of label indexing, label embedding, and input vectorizations on the \aol test set. For input embedding, we experiment with $\unigram{\prevQuery}$ and $\charngram{\prefix}$ where $\prevQuery, \prefix$ denote previous query and prefix respectively, $\unigram{\cdot}$ denotes unigram tfidf vectorization, $\phi_{\text{c-tfidf}}(\cdot)$ denotes character n-gram tfidf vectorization, and $\concat$ denotes concatenation. For vectorizing prefix or label (next query) text, we experiment with simple as well as position-weighted character n-gram tfidf vectorizer.}
  \label{tab:xmc_variants_all}
  \begin{tabular}{cccccccc}
    \toprule
    \multirow{2}{*}{Config ID} & \multirow{2}{*}{Input}   & Prefix/Label  & Label ($\ell$)     & Label      & \multirow{2}{*}{MRR}           & Latency (in ms)  & Training time \\
            & & Vectorizer    & Embedding & Indexing   &            & $p50, p99$       & (in mins) \\
    \midrule
    0   & $\prevQueryEmbed$       &   -                 & $\pifa$               & 2-means HC                        & .0585 & 4.4 , 5.6  & 24 \\ % 169
    1   & $\prevQueryEmbed$       &   Simple            & $\labelText$          & 2-means HC                        & .0653 & 4.5 , 5.5  & 24 \\ % 177
    \hline
    2   & $\prevQueryPrefixEmbed$ &   Simple            & $\pifa$               & 2-means HC                        & .1841 & 7.1 , 9.6  & 25 \\ % 170
    3   & $\prevQueryPrefixEmbed$ &   Simple            & $\labelText$          & 2-means HC                        & .2088 & 6.8 , 9.0  & 26 \\ % 176

    4   & $\prevQueryPrefixEmbed$ &   Pos-Wgtd          & $\pifa$               & 2-means HC                        & .2026 & 6.3 , 8.1  & 22 \\ % 171
    5   & $\prevQueryPrefixEmbed$ &   Pos-Wgtd          & $\labelText$          & 2-means HC                        & .2282 & 6.3 , 8.2  & 23 \\ % 172
    \hline
    6   & $\prevQueryPrefixEmbed$ &   Pos-Wgtd          & $\labelText$          & Hybrid Trie ($d_{trie}=1$)         & .2317 & 5.5 , 6.5  & 25 \\ % 192
    7   & $\prevQueryPrefixEmbed$ &   Pos-Wgtd          & $\labelText$          & Hybrid Trie ($d_{trie}=2$)         & .2315 & 5.4 , 6.5  & 26 \\ % 199
    8   & $\prevQueryPrefixEmbed$ &   Pos-Wgtd          & $\labelText$          & Hybrid Trie ($d_{trie}=3$)         & .2319 & 5.5 , 6.6  & 41 \\ % 206
    9   & $\prevQueryPrefixEmbed$ &   Pos-Wgtd          & -                     & Trie ($d_{trie}=16$)              & .2220 & 4.9 , 5.5  & 140 \\ % 175
    10  & $\prevQueryPrefixEmbed$ &   Simple            & -                      & Trie ($d_{trie}=16$)             & .2310 & 5.1 , 6.6  & 137 \\ % 174
    \multirow{2}{*}{11}   & \multirow{2}{*}{$\prevQueryPrefixEmbed$} &   \multirow{2}{*}{Pos-Wgtd}     & \multirow{2}{*}{$\labelText$}         &  2-means HC    & \multirow{2}{*}{.2291} & \multirow{2}{*}{6.0 , 7.6}  & \multirow{2}{*}{24} \\ % 185
        &               &                &                    & w/ \textsc{MLC} ($d_{mlc}=5$)    &       &             &     \\    
  \bottomrule
\end{tabular}
\end{table*}

\subsection{Comparing \our Models}
Table~\ref{tab:xmc_variants_all} shows a comparison of different \our models on 
\aol for different choices of input, vectorizers, and label indexing 
components. The baseline \our model (config-0 in Table~\ref{tab:xmc_variants_all}) 
which uses only the previous query as input with 
spherical 2-means hierarchical clustering based indexing corresponds to using \pecos out-of-the-box 
and performs poorly. Using both the prefix and previous query as part of the input to the ranking and matching 
models (config-2) yields a 3.1$\times$ improvement in MRR over using \pecos out-of-the-box. 
For all indexing methods, label 
embeddings obtained using label text tfidf vectors perform better than those 
obtained using the PIFA strategy.
The position weighted character n-gram tfidf 
vectorizer applied to the prefix and label yields up to 10\% improvement
over simple character n-gram tfidf vectorizers. 
Adding \emph{must-link} constraints to 2-means hierarchical clustering for indexing yields a small 
improvement when also using position weighted tfidf vectorizers. This is apparently 
because many \emph{must-link} constraints are already satisfied as a 
result of clustering with embeddings from position weighted tfidf vectorizers.
The best MRR is obtained by a \our model 
that indexes labels using a trie index. However, this model takes an order of 
magnitude longer to train as compared to hierarchical clustering based indices 
because of the imbalanced structure of the trie index. 
The hybrid trie index structures yield a small 
improvement over 2-means hierarchical clustering, and perform close to a pure 
trie based index while having much lower training time. 
\begin{table}
  \caption{Mean Reciprocal Rank (MRR), \RRBLEU scores and inference latency (in ms) of the top performing \our model ( config-8 from Table~\ref{tab:xmc_variants_all}) and baseline models on entire \aol test set, and on test data points for which next query is seen during training.}
  \label{tab:xmc_vs_baselines_all}
  \small
  \begin{tabular}{l|ccc|cc}
    \toprule
    \multicolumn{1}{c}{\multirow{4}{*}{Model}}          & \multicolumn{3}{|c}{\multirow{3}{*}{All Test Data}}     & \multicolumn{2}{|c}{Test Data} \\
                                &                       &                               &                         & \multicolumn{2}{|c}{w/ Seen} \\
                                &                       &                               &                         & \multicolumn{2}{|c}{Next Queries} \\
                                & \multirow{2}{*}{MRR}  &  \multirow{2}{*}{\RRBLEU}     & Latency                 & \multirow{2}{*}{MRR}      &   \multirow{2}{*}{\RRBLEU}\\
                                &                       &                               & $p50, p99$              &                           &   \\
    \midrule
     \our c-8                   & .231                  & .102                          & \textbf{5.5 , 6.5}      & \textbf{.572}             &  \textbf{.177} \\
     seq2seq-\GRU               & \textbf{.380}         & \textbf{.211}                 & 15.2, 22.1              & .544                      &  .165 \\ 
     \mfq                       & .225                  & .109                          & \textbf{0.01, 0.01}     & .556                      &  .147 \\
     \mfq +  seq2seq-\GRU       & .173                  & .089                          & 7.8, 9.1                & .426                      &  .125 \\
  \bottomrule
\end{tabular}
\end{table}

\subsection{Comparison with other baselines}
\begin{table*}
  \caption{Examples of nonsensical query suggestions generated by a seq2seq-GRU model for three previous query ($\prevQuery$), prefix ($\prefix$) pairs with grouth-truth next query ($\nextQuery$) from \aol test set. Prefix $\prefix$  of next query $\nextQuery$ is shown in bold.} 
  \label{tab:qualtitative}
  \footnotesize
  \begin{tabular}{clp{13cm}}
    \toprule
                            & Examples                  & \multicolumn{1}{c}{Auto-complete suggestions using seq2seq-GRU model}\\
    \midrule
    
    $\prevQuery$            & tortilla whole wheat nutrition facts  & \multirow{2}{\hsize}{\textbf{law of m}aking nutrition facts, \textbf{law of m}eals, \textbf{law of m}cdonalds, \textbf{law of m}ushroom facts, \textbf{law of m}arijuana, \textbf{law of m}ushroom} \\
    $\prefix, \nextQuery$   & \textbf{\emph{law of m}}otion & \\
     \hline
    $\prevQuery$            & ebay  & \multirow{2}{\hsize}{\textbf{cabbage machine} recipes, \textbf{cabbage machine} diet, \textbf{cabbage machine} shop, \textbf{cabbage machine} recipe, \textbf{cabbage machine} parts}\\
    $\prefix,\nextQuery$    & \textbf{\emph{cabbage machine}} cutter   &  \\
     \hline
    $\prevQuery$            & prepare your own wieght loss diet     &  \multirow{2}{\hsize}{\textbf{legal s}tudies, \textbf{legal s}tudies diet, \textbf{legal s}tudies wieght loss diet, \textbf{legal s}urveys, \textbf{legal s}teroids} \\
    $\prefix,\nextQuery$    & \textbf{\emph{legal s}}teriods       &  \\
  \bottomrule
\end{tabular}
\end{table*}

\begin{figure}
  \centering
  \includegraphics[width=\linewidth]{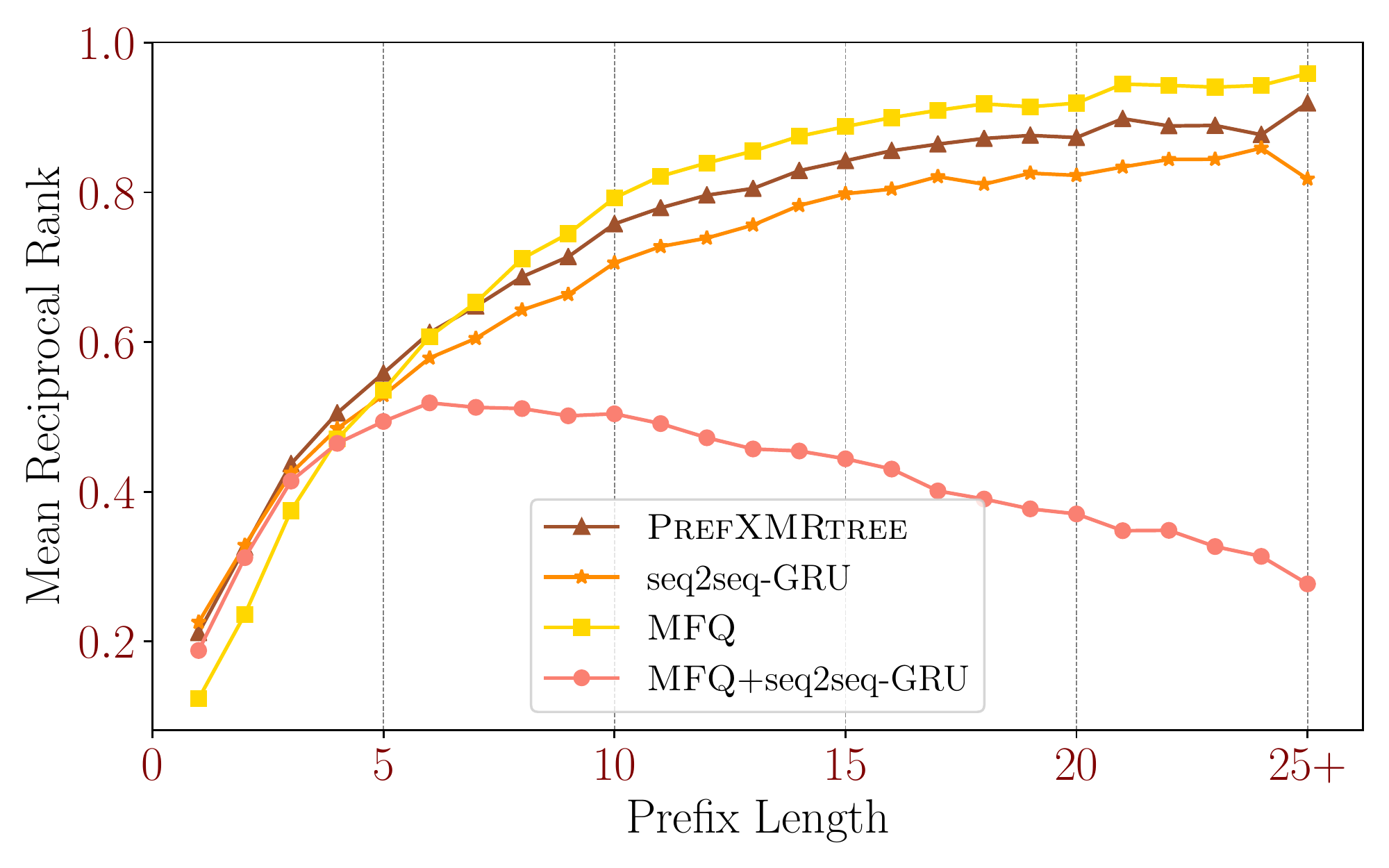}
  \caption{MRR of \our (c-8 in Table~\ref{tab:xmc_variants_all}) and baselines for different prefix lengths on 
  \aol test set where next query is seen at train time. }
  \label{fig:xmc_vs_baseline_pref}
\end{figure}

Table~\ref{tab:xmc_vs_baselines_all} shows a comparison of the best performing 
\our model (config-8 from Table~\ref{tab:xmc_variants_all}) with other baselines. 
Of the models that have latencies suitable for realtime systems with a
latency budget of the order of few milliseconds, \our outperforms 
other baselines such as \mfq, and \mfqplus.
Consistent with previous work~\cite{wang2020efficient}, \mfq 
outperforms \mfqplus which re-ranks
wrt likelihood conditioned on previous query using a seq2seq-GRU model (except for shorter prefixes as in Fig.~\ref{fig:xmc_vs_baseline_pref}).
seq2seq-GRU outperforms \our 
in terms on MRR and \RRBLEU scores but has a \emph{high} median and 
worst case inference latencies despite running on \emph{GPUs}, 
thus rendering it unsuitable for realtime systems. 
Moreover, as shown in Table~\ref{tab:qualtitative}, generative models like seq2seq-GRU
can generate  nonsensical auto-complete suggestions which could be undesirable in business
critical applications.

The primary reason for lower scores of \our models as compared to generative models 
is the finite label space derived from the training dataset.
For \aol, the training data contains 5.5MM unique labels (next queries) which
cover about 41\% of the test data point labels. Hence, for the remaining test data 
points, \our models and \mfq cannot generate correct suggestions, while  
seq2seq models are sometimes able to. The performance of \our can be improved 
by augmenting the label space to improve coverage 
and/or using a partial prefix for generating suggestions~\cite{mitra2015query, maxwell2017large} 
but we leave that to future work
and compare all the models on test data restricted to cases where the ground-truth label
(next query) is present in training data, and analyse the effect of prefix
length and label frequency on performance of all models.

Table~\ref{tab:xmc_vs_baselines_all} shows the performance of the best performing \our model
(config-8 from Table~\ref{tab:xmc_variants_all}) and baselines on the restricted test data. 
\our outperforms all the 
baselines in terms of MRR and \RRBLEU scores. 
\mfq, which corresponds to the maximum likelihood estimator for 
$P$(next query | prefix)~\cite{wang2018realtime}, serves as a strong baseline.
Note that, to its advantage, \mfq has direct access to a fixed set of labels (next queries) 
that exactly match the prefix whereas \our attempts to find and rank relevant labels 
from the \emph{entire} output space containing millions of labels.

Figure~\ref{fig:xmc_vs_baseline_pref} shows MRR for all models
for different prefix lengths on the restricted test data.
For all models, MRR improves with prefix length, matching 
the intuition that it becomes easier to 
predict the next query when the model has access to a larger prefix of it. 
Session-aware models such as \our, seq2seq-GRU and \mfqplus outperform \mfq for 
prefixes up to length 6 indicating that the context information from previous 
queries is most useful when the user has typed just a few characters of the 
next query. \our offers 71\%, 38\%, 17\%, 7.3\%, 4.3\%, 1\% improvement over \mfq 
for prefix lengths 1,2,3,4,5 and 6 respectively, and has the best inference latency among
session-aware models.

\begin{figure}
  \centering
  \includegraphics[width=\linewidth]{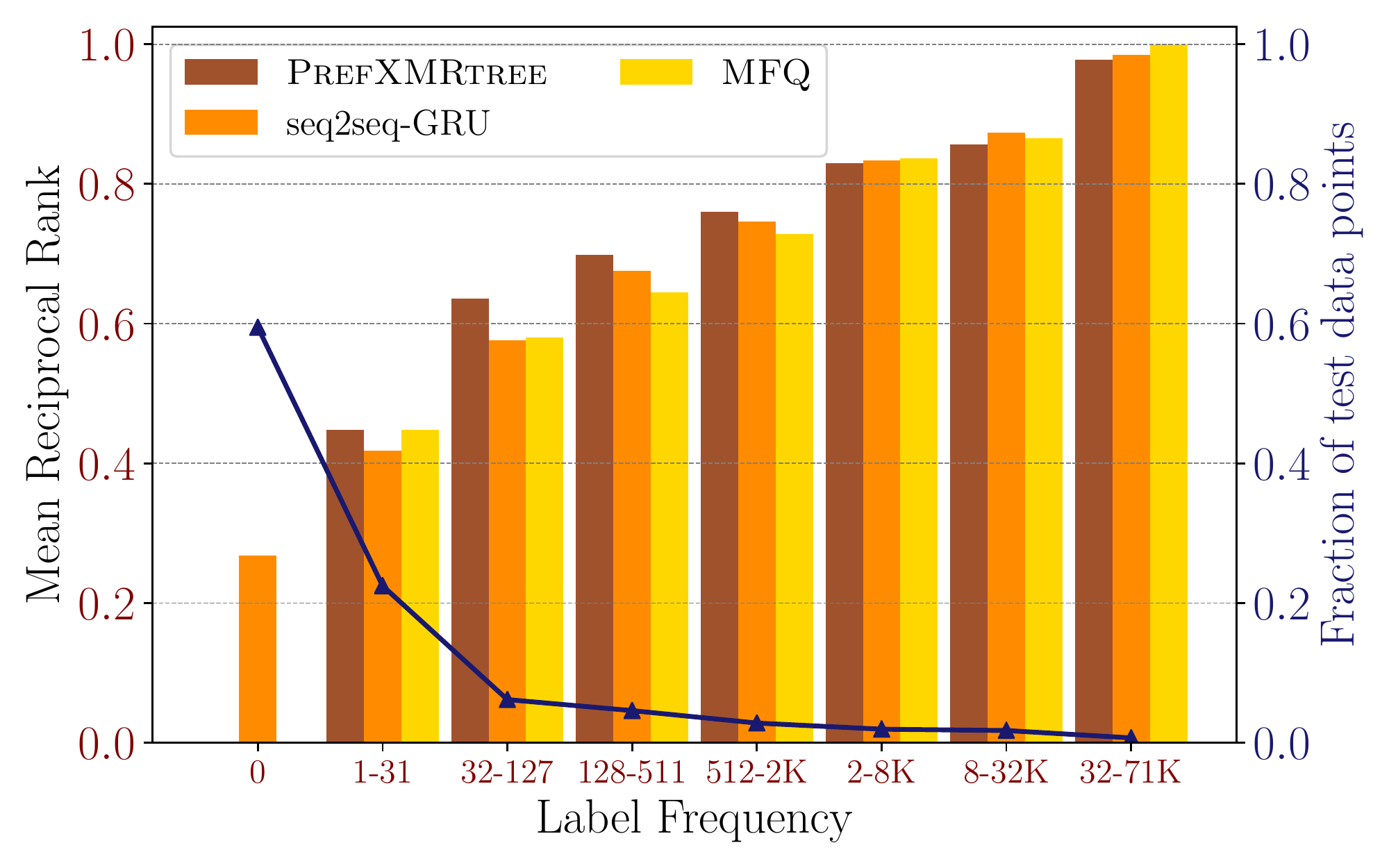}
  \caption{MRR of \our (c-8 in Table~\ref{tab:xmc_variants_all}) and baselines for different label frequency ranges on \aol test set where next query is seen at train time. Line plot on secondary y-axis shows the fraction of
 test data having labels with frequency in a given frequency range.}
  \label{fig:xmc_vs_baseline_freq}
\end{figure}

Figure~\ref{fig:xmc_vs_baseline_freq} shows the performance of all models 
bucketed by label frequency on the restricted test data 
and the fraction of test data with the label in a particular label frequency bucket.
Generally, the performance of all models improves with label frequency. 
Note that a significant fraction of test data contains queries with frequency less than 2,000,
and on average, \our outperforms \mfq for labels with frequency up to 2,000 while \mfq performs 
better for labels with frequency higher than 2,000. 

\subsection{Results on dataset from an Online Shopping Store}
Table~\ref{tab:xmc_vs_baselines_amazon} shows the percentage change in MRR for different 
models over \mfq baseline on search logs from an Online Shopping Store. 
We omit \mfqplus baseline because of its poor performance on \aol. 
seq2seq-GRU performs the best on entire test data but despite using GPUs, its inference 
latency and throughput are \emph{unsuitable} for real-time systems where the latency budget is 
of the order of milliseconds. When the next query is present in training data, 
\our offers a significant 17\% improvement over seq2seq-GRU, 
and 1\% improvement over \mfq. A further breakdown of MRR based on prefix 
length shows that \our offers up to 143\% improvement for prefix length 1, and 
24\%  improvement over \mfq for small prefixes up to length 8. 

A version of \our was deployed online on the search bar of the Online Shopping Store 
as part of an A/B test, where it yielded a statistically significant improvement of 2.81\% over the production 
system in terms of \qac suggestion acceptance rate. For reasons discussed
before such as high inference latency despite running on the more expensive GPUs and the 
potential risk of generating nonsensical auto-complete suggestions, we could not
deploy seq2seq-GRU model as part of the A/B test.

\begin{table}
  \caption{Percentage change in MRR for \our (c-5 from Table~\ref{tab:xmc_variants_all}) and seq2seq-GRU over \mfq baseline on test set from search logs of an online shopping store. 
  }
  \label{tab:xmc_vs_baselines_amazon}
  \begin{tabular}{l|c|cccc}
    \toprule
   \multicolumn{1}{c}{\multirow{3}{*}{Model}}  & \multicolumn{1}{|c|}{\multirow{3}{*}{\parbox{1cm}{All Test   \text{  }  Data}}}    & \multicolumn{4}{c}{Test Data w/ Seen Next Queries} \\\cline{3-6}

                              &         & \multicolumn{4}{c}{Prefix Length}  \\ 
                              &             &  All    & 1           & 2-8         & $\geq$ 9\\
    \midrule     
     seq2seq-\GRU            & \textbf{+37} & -12.5       &   +136.8          &   +6.0            &     -24.7     \\ % 134
     \our                    & +1           & \textbf{+1} &   \textbf{+143.2} &   \textbf{+24}  &     -14.3     \\ % 117
  \bottomrule
\end{tabular}
\vspace{-0.5cm}
\end{table}

\section{Conclusion}
In this paper, we propose \our, a tree-based \xmr model for  
the task of session-aware query auto-completion under stringent 
latency requirements. We propose and benchmark several novel label 
indexing and embedding schemes which make the \our models more suitable 
than an out-of-the-box \xmr model for retrieving relevant suggestions  
conditioned on the previous query while adhering to the user's prefix.
Overall, the proposed \xmr models provide a better combination of statistical 
performance and inference latency as compared to other baselines while 
specifically improving suggestions for smaller prefix lengths, where the
contextual information from previous queries has the most utility.

In the future, we plan to improve the performance of \our models 
by exploring label space augmentation techniques and using partial 
prefixes for generating auto-complete 
suggestions~\cite{mitra2015query, maxwell2017large}. We also plan to use 
dense embeddings for representing contextual information such as previous 
queries, the user-profile, or time to further improve the performance of 
\our models for the task of query auto-completion.

\bibliographystyle{ACM-Reference-Format}
\bibliography{kdd_2021}


\begin{thebibliography}{46}

%%% the Web address.

\ifx \showCODEN    \undefined \def \showCODEN     #1{\unskip}     \fi
\ifx \showDOI      \undefined \def \showDOI       #1{#1}\fi
\ifx \showISBNx    \undefined \def \showISBNx     #1{\unskip}     \fi
\ifx \showISBNxiii \undefined \def \showISBNxiii  #1{\unskip}     \fi
\ifx \showISSN     \undefined \def \showISSN      #1{\unskip}     \fi
\ifx \showLCCN     \undefined \def \showLCCN      #1{\unskip}     \fi
\ifx \shownote     \undefined \def \shownote      #1{#1}          \fi
\ifx \showarticletitle \undefined \def \showarticletitle #1{#1}   \fi
\ifx \showURL      \undefined \def \showURL       {\relax}        \fi
% invisible to TeX
\providecommand\bibfield[2]{#2}
\providecommand\bibinfo[2]{#2}
\providecommand\natexlab[1]{#1}
\providecommand\showeprint[2][]{arXiv:#2}

\bibitem[\protect\citeauthoryear{Babbar and Sch{\"o}lkopf}{Babbar and
  Sch{\"o}lkopf}{2017}]%
        {babbar2017dismec}
\bibfield{author}{\bibinfo{person}{Rohit Babbar} {and}
  \bibinfo{person}{Bernhard Sch{\"o}lkopf}.} \bibinfo{year}{2017}\natexlab{}.
\newblock \showarticletitle{Dismec: Distributed sparse machines for extreme
  multi-label classification}. In \bibinfo{booktitle}{\emph{Proceedings of the
  10th ACM International Conference on Web Search and Data Mining}}.
  \bibinfo{pages}{721--729}.
\newblock


\bibitem[\protect\citeauthoryear{Bar-Yossef and Kraus}{Bar-Yossef and
  Kraus}{2011}]%
        {bar2011context}
\bibfield{author}{\bibinfo{person}{Ziv Bar-Yossef} {and} \bibinfo{person}{Naama
  Kraus}.} \bibinfo{year}{2011}\natexlab{}.
\newblock \showarticletitle{Context-sensitive query auto-completion}. In
  \bibinfo{booktitle}{\emph{Proceedings of the 20th International Conference on
  World Wide Web}}. \bibinfo{pages}{107--116}.
\newblock


\bibitem[\protect\citeauthoryear{Bhatia, Jain, Kar, Varma, and Jain}{Bhatia
  et~al\mbox{.}}{2015}]%
        {bhatia2015sparse}
\bibfield{author}{\bibinfo{person}{Kush Bhatia}, \bibinfo{person}{Himanshu
  Jain}, \bibinfo{person}{Purushottam Kar}, \bibinfo{person}{Manik Varma},
  {and} \bibinfo{person}{Prateek Jain}.} \bibinfo{year}{2015}\natexlab{}.
\newblock \showarticletitle{Sparse local embeddings for extreme multi-label
  classification}. In \bibinfo{booktitle}{\emph{Advances in Neural Information
  Processing Systems}}. \bibinfo{pages}{730--738}.
\newblock


\bibitem[\protect\citeauthoryear{Cai and de~Rijke}{Cai and de~Rijke}{2016}]%
        {cai2016query}
\bibfield{author}{\bibinfo{person}{Fei Cai} {and} \bibinfo{person}{Maarten de
  Rijke}.} \bibinfo{year}{2016}\natexlab{}.
\newblock \bibinfo{booktitle}{\emph{A Survey of Query Auto Completion in
  Information Retrieval}}.
\newblock \bibinfo{publisher}{Now Publishers Inc.}
\newblock
\showISBNx{168083200X}


\bibitem[\protect\citeauthoryear{Cai, Liang, and De~Rijke}{Cai
  et~al\mbox{.}}{2014}]%
        {cai2014time}
\bibfield{author}{\bibinfo{person}{Fei Cai}, \bibinfo{person}{Shangsong Liang},
  {and} \bibinfo{person}{Maarten De~Rijke}.} \bibinfo{year}{2014}\natexlab{}.
\newblock \showarticletitle{Time-sensitive personalized query auto-completion}.
  In \bibinfo{booktitle}{\emph{Proceedings of the 23rd ACM International
  Conference on Information and Knowledge Management}}.
  \bibinfo{pages}{1599--1608}.
\newblock


\bibitem[\protect\citeauthoryear{Chen and Cherry}{Chen and Cherry}{2014}]%
        {chen2014systematic}
\bibfield{author}{\bibinfo{person}{Boxing Chen} {and} \bibinfo{person}{Colin
  Cherry}.} \bibinfo{year}{2014}\natexlab{}.
\newblock \showarticletitle{A systematic comparison of smoothing techniques for
  sentence-level bleu}. In \bibinfo{booktitle}{\emph{Proceedings of the Ninth
  Workshop on Statistical Machine Translation}}. \bibinfo{pages}{362--367}.
\newblock


\bibitem[\protect\citeauthoryear{Chen and Lin}{Chen and Lin}{2012}]%
        {chen2012feature}
\bibfield{author}{\bibinfo{person}{Yao-Nan Chen} {and}
  \bibinfo{person}{Hsuan-Tien Lin}.} \bibinfo{year}{2012}\natexlab{}.
\newblock \showarticletitle{Feature-aware label space dimension reduction for
  multi-label classification}. In \bibinfo{booktitle}{\emph{Advances in Neural
  Information Processing Systems}}. \bibinfo{pages}{1529--1537}.
\newblock


\bibitem[\protect\citeauthoryear{Cho, Van~Merri{\"e}nboer, Gulcehre, Bahdanau,
  Bougares, Schwenk, and Bengio}{Cho et~al\mbox{.}}{2014}]%
        {cho2014learning}
\bibfield{author}{\bibinfo{person}{Kyunghyun Cho}, \bibinfo{person}{Bart
  Van~Merri{\"e}nboer}, \bibinfo{person}{Caglar Gulcehre},
  \bibinfo{person}{Dzmitry Bahdanau}, \bibinfo{person}{Fethi Bougares},
  \bibinfo{person}{Holger Schwenk}, {and} \bibinfo{person}{Yoshua Bengio}.}
  \bibinfo{year}{2014}\natexlab{}.
\newblock \showarticletitle{Learning Phrase Representations using {RNN}
  Encoder{--}Decoder for Statistical Machine Translation}. In
  \bibinfo{booktitle}{\emph{Empirical Methods in Natural Language Processing}}.
\newblock


\bibitem[\protect\citeauthoryear{Cisse, Usunier, Artieres, and Gallinari}{Cisse
  et~al\mbox{.}}{2013}]%
        {cisse2013robust}
\bibfield{author}{\bibinfo{person}{Moustapha~M Cisse}, \bibinfo{person}{Nicolas
  Usunier}, \bibinfo{person}{Thierry Artieres}, {and} \bibinfo{person}{Patrick
  Gallinari}.} \bibinfo{year}{2013}\natexlab{}.
\newblock \showarticletitle{Robust bloom filters for large multilabel
  classification tasks}. In \bibinfo{booktitle}{\emph{Advances in Neural
  Information Processing Systems}}. \bibinfo{pages}{1851--1859}.
\newblock


\bibitem[\protect\citeauthoryear{Dehghani, Rothe, Alfonseca, and
  Fleury}{Dehghani et~al\mbox{.}}{2017}]%
        {dehghani2017learning}
\bibfield{author}{\bibinfo{person}{Mostafa Dehghani}, \bibinfo{person}{Sascha
  Rothe}, \bibinfo{person}{Enrique Alfonseca}, {and} \bibinfo{person}{Pascal
  Fleury}.} \bibinfo{year}{2017}\natexlab{}.
\newblock \showarticletitle{Learning to attend, copy, and generate for
  session-based query suggestion}. In \bibinfo{booktitle}{\emph{Proceedings of
  the 26th ACM International Conference on Information and Knowledge
  Management}}. \bibinfo{pages}{1747--1756}.
\newblock


\bibitem[\protect\citeauthoryear{Devlin, Chang, Lee, and Toutanova}{Devlin
  et~al\mbox{.}}{2019}]%
        {devlin2018bert}
\bibfield{author}{\bibinfo{person}{Jacob Devlin}, \bibinfo{person}{Ming-Wei
  Chang}, \bibinfo{person}{Kenton Lee}, {and} \bibinfo{person}{Kristina
  Toutanova}.} \bibinfo{year}{2019}\natexlab{}.
\newblock \showarticletitle{{BERT}: Pre-training of Deep Bidirectional
  Transformers for Language Understanding}.
\newblock  (\bibinfo{year}{2019}), \bibinfo{pages}{4171--4186}.
\newblock


\bibitem[\protect\citeauthoryear{Fiorini and Lu}{Fiorini and Lu}{2018}]%
        {fiorini2018personalized}
\bibfield{author}{\bibinfo{person}{Nicolas Fiorini} {and}
  \bibinfo{person}{Zhiyong Lu}.} \bibinfo{year}{2018}\natexlab{}.
\newblock \showarticletitle{Personalized neural language models for real-world
  query auto completion}. In \bibinfo{booktitle}{\emph{North American Chapter
  of the Association for Computational Linguistics: Human Language
  Technologies}}. \bibinfo{pages}{208--215}.
\newblock


\bibitem[\protect\citeauthoryear{Hsu and Ottaviano}{Hsu and Ottaviano}{2013}]%
        {hsu2013space}
\bibfield{author}{\bibinfo{person}{Bo-June Hsu} {and} \bibinfo{person}{Giuseppe
  Ottaviano}.} \bibinfo{year}{2013}\natexlab{}.
\newblock \showarticletitle{Space-efficient data structures for top-k
  completion}. In \bibinfo{booktitle}{\emph{Proceedings of the 22nd
  International Conference on World Wide Web}}. \bibinfo{pages}{583--594}.
\newblock


\bibitem[\protect\citeauthoryear{Jaech and Ostendorf}{Jaech and
  Ostendorf}{2018}]%
        {jaech2018personalized}
\bibfield{author}{\bibinfo{person}{Aaron Jaech} {and} \bibinfo{person}{Mari
  Ostendorf}.} \bibinfo{year}{2018}\natexlab{}.
\newblock \showarticletitle{Personalized Language Model for Query
  Auto-Completion}. In \bibinfo{booktitle}{\emph{Proceedings of the 56th Annual
  Meeting of the Association for Computational Linguistics}}.
  \bibinfo{pages}{700--705}.
\newblock


\bibitem[\protect\citeauthoryear{Jain, Balasubramanian, Chunduri, and
  Varma}{Jain et~al\mbox{.}}{2019}]%
        {jain2019slice}
\bibfield{author}{\bibinfo{person}{Himanshu Jain}, \bibinfo{person}{Venkatesh
  Balasubramanian}, \bibinfo{person}{Bhanu Chunduri}, {and}
  \bibinfo{person}{Manik Varma}.} \bibinfo{year}{2019}\natexlab{}.
\newblock \showarticletitle{Slice: Scalable linear extreme classifiers trained
  on 100 million labels for related searches}. In
  \bibinfo{booktitle}{\emph{Proceedings of the 12th ACM International
  Conference on Web Search and Data Mining}}. \bibinfo{pages}{528--536}.
\newblock


\bibitem[\protect\citeauthoryear{Jasinska, Dembczynski, Busa-Fekete,
  Pfannschmidt, Klerx, and Hullermeier}{Jasinska et~al\mbox{.}}{2016}]%
        {jasinska2016extreme}
\bibfield{author}{\bibinfo{person}{Kalina Jasinska}, \bibinfo{person}{Krzysztof
  Dembczynski}, \bibinfo{person}{R{\'o}bert Busa-Fekete},
  \bibinfo{person}{Karlson Pfannschmidt}, \bibinfo{person}{Timo Klerx}, {and}
  \bibinfo{person}{Eyke Hullermeier}.} \bibinfo{year}{2016}\natexlab{}.
\newblock \showarticletitle{Extreme F-measure maximization using sparse
  probability estimates}. In \bibinfo{booktitle}{\emph{International Conference
  on Machine Learning}}. \bibinfo{pages}{1435--1444}.
\newblock


\bibitem[\protect\citeauthoryear{Jiang, Chen, Cai, and Chen}{Jiang
  et~al\mbox{.}}{2018}]%
        {jiang2018neural}
\bibfield{author}{\bibinfo{person}{Danyang Jiang}, \bibinfo{person}{Wanyu
  Chen}, \bibinfo{person}{Fei Cai}, {and} \bibinfo{person}{Honghui Chen}.}
  \bibinfo{year}{2018}\natexlab{}.
\newblock \showarticletitle{Neural Attentive Personalization Model for Query
  Auto-Completion}. In \bibinfo{booktitle}{\emph{IEEE 3rd Advanced Information
  Technology, Electronic and Automation Control Conference}}.
  \bibinfo{pages}{725--730}.
\newblock


\bibitem[\protect\citeauthoryear{Jiang, Ke, Chien, and Cheng}{Jiang
  et~al\mbox{.}}{2014}]%
        {jiang2014learning}
\bibfield{author}{\bibinfo{person}{Jyun-Yu Jiang}, \bibinfo{person}{Yen-Yu Ke},
  \bibinfo{person}{Pao-Yu Chien}, {and} \bibinfo{person}{Pu-Jen Cheng}.}
  \bibinfo{year}{2014}\natexlab{}.
\newblock \showarticletitle{Learning user reformulation behavior for query
  auto-completion}. In \bibinfo{booktitle}{\emph{Proceedings of the 37th
  International ACM SIGIR Conference on Research and Development in Information
  Retrieval}}. \bibinfo{pages}{445--454}.
\newblock


\bibitem[\protect\citeauthoryear{Kastrinakis and Tzitzikas}{Kastrinakis and
  Tzitzikas}{2010}]%
        {kastrinakis2010advancing}
\bibfield{author}{\bibinfo{person}{Dimitrios Kastrinakis} {and}
  \bibinfo{person}{Yannis Tzitzikas}.} \bibinfo{year}{2010}\natexlab{}.
\newblock \showarticletitle{Advancing search query autocompletion services with
  more and better suggestions}. In \bibinfo{booktitle}{\emph{International
  Conference on Web Engineering}}. \bibinfo{pages}{35--49}.
\newblock


\bibitem[\protect\citeauthoryear{Khandagale, Xiao, and Babbar}{Khandagale
  et~al\mbox{.}}{2020}]%
        {khandagale2020bonsai}
\bibfield{author}{\bibinfo{person}{Sujay Khandagale}, \bibinfo{person}{Han
  Xiao}, {and} \bibinfo{person}{Rohit Babbar}.}
  \bibinfo{year}{2020}\natexlab{}.
\newblock \showarticletitle{Bonsai: diverse and shallow trees for extreme
  multi-label classification}.
\newblock \bibinfo{journal}{\emph{Machine Learning}} (\bibinfo{year}{2020}),
  \bibinfo{pages}{1--21}.
\newblock


\bibitem[\protect\citeauthoryear{Kingma and Ba}{Kingma and Ba}{2015}]%
        {kingma2014adam}
\bibfield{author}{\bibinfo{person}{Diederik~P Kingma} {and}
  \bibinfo{person}{Jimmy Ba}.} \bibinfo{year}{2015}\natexlab{}.
\newblock \showarticletitle{Adam: A method for stochastic optimization}.
\newblock \bibinfo{journal}{\emph{International Conference on Learning
  Representations}} (\bibinfo{year}{2015}).
\newblock


\bibitem[\protect\citeauthoryear{Kudo and Richardson}{Kudo and
  Richardson}{2018}]%
        {kudo2018sentencepiece}
\bibfield{author}{\bibinfo{person}{Taku Kudo} {and} \bibinfo{person}{John
  Richardson}.} \bibinfo{year}{2018}\natexlab{}.
\newblock \showarticletitle{{S}entence{P}iece: A simple and language
  independent subword tokenizer and detokenizer for Neural Text Processing}. In
  \bibinfo{booktitle}{\emph{Empirical Methods in Natural Language Processing}}.
\newblock


\bibitem[\protect\citeauthoryear{Lin, Ding, Hu, and Wang}{Lin
  et~al\mbox{.}}{2014}]%
        {lin2014multi}
\bibfield{author}{\bibinfo{person}{Zijia Lin}, \bibinfo{person}{Guiguang Ding},
  \bibinfo{person}{Mingqing Hu}, {and} \bibinfo{person}{Jianmin Wang}.}
  \bibinfo{year}{2014}\natexlab{}.
\newblock \showarticletitle{Multi-label classification via feature-aware
  implicit label space encoding}. In \bibinfo{booktitle}{\emph{International
  Conference on Machine Learning}}. \bibinfo{pages}{325--333}.
\newblock


\bibitem[\protect\citeauthoryear{Malkov and Yashunin}{Malkov and
  Yashunin}{2018}]%
        {malkov2018efficient}
\bibfield{author}{\bibinfo{person}{Yury~A Malkov} {and}
  \bibinfo{person}{Dmitry~A Yashunin}.} \bibinfo{year}{2018}\natexlab{}.
\newblock \showarticletitle{Efficient and robust approximate nearest neighbor
  search using hierarchical navigable small world graphs}.
\newblock \bibinfo{journal}{\emph{IEEE transactions on pattern analysis and
  machine intelligence}} (\bibinfo{year}{2018}).
\newblock


\bibitem[\protect\citeauthoryear{Maxwell, Bailey, and Hawking}{Maxwell
  et~al\mbox{.}}{2017}]%
        {maxwell2017large}
\bibfield{author}{\bibinfo{person}{David Maxwell}, \bibinfo{person}{Peter
  Bailey}, {and} \bibinfo{person}{David Hawking}.}
  \bibinfo{year}{2017}\natexlab{}.
\newblock \showarticletitle{Large-scale generative query autocompletion}. In
  \bibinfo{booktitle}{\emph{Proceedings of the 22nd Australasian Document
  Computing Symposium}}. \bibinfo{pages}{1--8}.
\newblock


\bibitem[\protect\citeauthoryear{Mineiro and Karampatziakis}{Mineiro and
  Karampatziakis}{2015}]%
        {mineiro2015fast}
\bibfield{author}{\bibinfo{person}{Paul Mineiro} {and} \bibinfo{person}{Nikos
  Karampatziakis}.} \bibinfo{year}{2015}\natexlab{}.
\newblock \showarticletitle{Fast label embeddings for extremely large output
  spaces}.
\newblock \bibinfo{journal}{\emph{arXiv preprint arXiv:1503.08873}}
  (\bibinfo{year}{2015}).
\newblock


\bibitem[\protect\citeauthoryear{Mitra and Craswell}{Mitra and
  Craswell}{2015}]%
        {mitra2015query}
\bibfield{author}{\bibinfo{person}{Bhaskar Mitra} {and} \bibinfo{person}{Nick
  Craswell}.} \bibinfo{year}{2015}\natexlab{}.
\newblock \showarticletitle{Query auto-completion for rare prefixes}. In
  \bibinfo{booktitle}{\emph{Proceedings of the 24th ACM International
  Conference on Information and Knowledge Management}}.
  \bibinfo{pages}{1755--1758}.
\newblock


\bibitem[\protect\citeauthoryear{Mustar, Lamprier, and Piwowarski}{Mustar
  et~al\mbox{.}}{2020}]%
        {mustarusing}
\bibfield{author}{\bibinfo{person}{Agn{\`e}s Mustar}, \bibinfo{person}{Sylvain
  Lamprier}, {and} \bibinfo{person}{Benjamin Piwowarski}.}
  \bibinfo{year}{2020}\natexlab{}.
\newblock \showarticletitle{Using BERT and BART for Query Suggestion}. In
  \bibinfo{booktitle}{\emph{Joint Conference of the Information Retrieval
  Communities in Europe}}.
\newblock


\bibitem[\protect\citeauthoryear{Papineni, Roukos, Ward, and Zhu}{Papineni
  et~al\mbox{.}}{2002}]%
        {papineni2002bleu}
\bibfield{author}{\bibinfo{person}{Kishore Papineni}, \bibinfo{person}{Salim
  Roukos}, \bibinfo{person}{Todd Ward}, {and} \bibinfo{person}{Wei-Jing Zhu}.}
  \bibinfo{year}{2002}\natexlab{}.
\newblock \showarticletitle{BLEU: a method for automatic evaluation of machine
  translation}. In \bibinfo{booktitle}{\emph{Proceedings of the 40th Annual
  Meeting of the Association for Computational Linguistics}}.
  \bibinfo{pages}{311--318}.
\newblock


\bibitem[\protect\citeauthoryear{Park and Chiba}{Park and Chiba}{2017}]%
        {park2017neural}
\bibfield{author}{\bibinfo{person}{Dae~Hoon Park} {and} \bibinfo{person}{Rikio
  Chiba}.} \bibinfo{year}{2017}\natexlab{}.
\newblock \showarticletitle{A neural language model for query auto-completion}.
  In \bibinfo{booktitle}{\emph{Proceedings of the 40th International ACM SIGIR
  Conference on Research and Development in Information Retrieval}}.
  \bibinfo{pages}{1189--1192}.
\newblock


\bibitem[\protect\citeauthoryear{Pass, Chowdhury, and Torgeson}{Pass
  et~al\mbox{.}}{2006}]%
        {pass2006picture}
\bibfield{author}{\bibinfo{person}{Greg Pass}, \bibinfo{person}{Abdur
  Chowdhury}, {and} \bibinfo{person}{Cayley Torgeson}.}
  \bibinfo{year}{2006}\natexlab{}.
\newblock \showarticletitle{A picture of search}. In
  \bibinfo{booktitle}{\emph{Proceedings of the 1st International Conference on
  Scalable Information Systems}}.
\newblock


\bibitem[\protect\citeauthoryear{Pedregosa, Varoquaux, Gramfort, Michel,
  Thirion, Grisel, Blondel, Prettenhofer, Weiss, Dubourg,
  et~al\mbox{.}}{Pedregosa et~al\mbox{.}}{2011}]%
        {pedregosa2011scikit}
\bibfield{author}{\bibinfo{person}{Fabian Pedregosa}, \bibinfo{person}{Ga{\"e}l
  Varoquaux}, \bibinfo{person}{Alexandre Gramfort}, \bibinfo{person}{Vincent
  Michel}, \bibinfo{person}{Bertrand Thirion}, \bibinfo{person}{Olivier
  Grisel}, \bibinfo{person}{Mathieu Blondel}, \bibinfo{person}{Peter
  Prettenhofer}, \bibinfo{person}{Ron Weiss}, \bibinfo{person}{Vincent
  Dubourg}, {et~al\mbox{.}}} \bibinfo{year}{2011}\natexlab{}.
\newblock \showarticletitle{Scikit-learn: Machine learning in Python}.
\newblock \bibinfo{journal}{\emph{the Journal of Machine Learning Research}}
  \bibinfo{volume}{12} (\bibinfo{year}{2011}), \bibinfo{pages}{2825--2830}.
\newblock


\bibitem[\protect\citeauthoryear{Prabhu, Kag, Harsola, Agrawal, and
  Varma}{Prabhu et~al\mbox{.}}{2018}]%
        {prabhu2018parabel}
\bibfield{author}{\bibinfo{person}{Yashoteja Prabhu}, \bibinfo{person}{Anil
  Kag}, \bibinfo{person}{Shrutendra Harsola}, \bibinfo{person}{Rahul Agrawal},
  {and} \bibinfo{person}{Manik Varma}.} \bibinfo{year}{2018}\natexlab{}.
\newblock \showarticletitle{Parabel: Partitioned label trees for extreme
  classification with application to dynamic search advertising}. In
  \bibinfo{booktitle}{\emph{Proceedings of the 27th International Conference on
  World Wide Web}}. \bibinfo{pages}{993--1002}.
\newblock


\bibitem[\protect\citeauthoryear{Shokouhi}{Shokouhi}{2013}]%
        {shokouhi2013learning}
\bibfield{author}{\bibinfo{person}{Milad Shokouhi}.}
  \bibinfo{year}{2013}\natexlab{}.
\newblock \showarticletitle{Learning to personalize query auto-completion}. In
  \bibinfo{booktitle}{\emph{Proceedings of the 36th International ACM SIGIR
  Conference on Research and Development in Information Retrieval}}.
  \bibinfo{pages}{103--112}.
\newblock


\bibitem[\protect\citeauthoryear{Song, Xiao, Wu, Wu, Zhang, Zhang, and
  Zhu}{Song et~al\mbox{.}}{2017}]%
        {song2017hierarchical}
\bibfield{author}{\bibinfo{person}{Jun Song}, \bibinfo{person}{Jun Xiao},
  \bibinfo{person}{Fei Wu}, \bibinfo{person}{Haishan Wu}, \bibinfo{person}{Tong
  Zhang}, \bibinfo{person}{Zhongfei~Mark Zhang}, {and} \bibinfo{person}{Wenwu
  Zhu}.} \bibinfo{year}{2017}\natexlab{}.
\newblock \showarticletitle{Hierarchical contextual attention recurrent neural
  network for map query suggestion}.
\newblock \bibinfo{journal}{\emph{IEEE Transactions on Knowledge and Data
  Engineering}} \bibinfo{volume}{29}, \bibinfo{number}{9}
  (\bibinfo{year}{2017}), \bibinfo{pages}{1888--1901}.
\newblock


\bibitem[\protect\citeauthoryear{Sordoni, Bengio, Vahabi, Lioma, Grue~Simonsen,
  and Nie}{Sordoni et~al\mbox{.}}{2015}]%
        {sordoni2015hierarchical}
\bibfield{author}{\bibinfo{person}{Alessandro Sordoni}, \bibinfo{person}{Yoshua
  Bengio}, \bibinfo{person}{Hossein Vahabi}, \bibinfo{person}{Christina Lioma},
  \bibinfo{person}{Jakob Grue~Simonsen}, {and} \bibinfo{person}{Jian-Yun Nie}.}
  \bibinfo{year}{2015}\natexlab{}.
\newblock \showarticletitle{A hierarchical recurrent encoder-decoder for
  generative context-aware query suggestion}. In
  \bibinfo{booktitle}{\emph{Proceedings of the 24th ACM International
  Conference on Information and Knowledge Management}}.
  \bibinfo{pages}{553--562}.
\newblock


\bibitem[\protect\citeauthoryear{Sparck~Jones}{Sparck~Jones}{1988}]%
        {jones1972statistical}
\bibfield{author}{\bibinfo{person}{Karen Sparck~Jones}.}
  \bibinfo{year}{1988}\natexlab{}.
\newblock \bibinfo{booktitle}{\emph{A Statistical Interpretation of Term
  Specificity and Its Application in Retrieval}}.
\newblock \bibinfo{pages}{132–142}.
\newblock


\bibitem[\protect\citeauthoryear{Wang, Zhang, Mohan, Dhillon, and Kolter}{Wang
  et~al\mbox{.}}{2018}]%
        {wang2018realtime}
\bibfield{author}{\bibinfo{person}{Po-Wei Wang}, \bibinfo{person}{Huan Zhang},
  \bibinfo{person}{Vijai Mohan}, \bibinfo{person}{Inderjit~S Dhillon}, {and}
  \bibinfo{person}{J~Zico Kolter}.} \bibinfo{year}{2018}\natexlab{}.
\newblock \showarticletitle{Realtime query completion via deep language
  models}. In \bibinfo{booktitle}{\emph{eCOM@ SIGIR}}.
\newblock


\bibitem[\protect\citeauthoryear{Wang, Guo, Gao, and Long}{Wang
  et~al\mbox{.}}{2020}]%
        {wang2020efficient}
\bibfield{author}{\bibinfo{person}{Sida Wang}, \bibinfo{person}{Weiwei Guo},
  \bibinfo{person}{Huiji Gao}, {and} \bibinfo{person}{Bo Long}.}
  \bibinfo{year}{2020}\natexlab{}.
\newblock \showarticletitle{Efficient Neural Query Auto Completion}. In
  \bibinfo{booktitle}{\emph{Proceedings of the 29th ACM International
  Conference on Information and Knowledge Management}}.
  \bibinfo{pages}{2797--2804}.
\newblock


\bibitem[\protect\citeauthoryear{Weston, Makadia, and Yee}{Weston
  et~al\mbox{.}}{2013}]%
        {weston2013label}
\bibfield{author}{\bibinfo{person}{Jason Weston}, \bibinfo{person}{Ameesh
  Makadia}, {and} \bibinfo{person}{Hector Yee}.}
  \bibinfo{year}{2013}\natexlab{}.
\newblock \showarticletitle{Label partitioning for sublinear ranking}. In
  \bibinfo{booktitle}{\emph{International Conference on Machine Learning}}.
  \bibinfo{pages}{181--189}.
\newblock


\bibitem[\protect\citeauthoryear{Wu, Burges, Svore, and Gao}{Wu
  et~al\mbox{.}}{2010}]%
        {wu2010adapting}
\bibfield{author}{\bibinfo{person}{Qiang Wu}, \bibinfo{person}{Christopher~JC
  Burges}, \bibinfo{person}{Krysta~M Svore}, {and} \bibinfo{person}{Jianfeng
  Gao}.} \bibinfo{year}{2010}\natexlab{}.
\newblock \showarticletitle{Adapting boosting for information retrieval
  measures}.
\newblock \bibinfo{journal}{\emph{Information Retrieval}} \bibinfo{volume}{13},
  \bibinfo{number}{3} (\bibinfo{year}{2010}), \bibinfo{pages}{254--270}.
\newblock


\bibitem[\protect\citeauthoryear{Xiao, Qin, Wang, Ishikawa, Tsuda, and
  Sadakane}{Xiao et~al\mbox{.}}{2013}]%
        {xiao2013efficient}
\bibfield{author}{\bibinfo{person}{Chuan Xiao}, \bibinfo{person}{Jianbin Qin},
  \bibinfo{person}{Wei Wang}, \bibinfo{person}{Yoshiharu Ishikawa},
  \bibinfo{person}{Koji Tsuda}, {and} \bibinfo{person}{Kunihiko Sadakane}.}
  \bibinfo{year}{2013}\natexlab{}.
\newblock \showarticletitle{Efficient error-tolerant query autocompletion}.
\newblock \bibinfo{journal}{\emph{Proceedings of the VLDB Endowment}}
  \bibinfo{volume}{6}, \bibinfo{number}{6} (\bibinfo{year}{2013}),
  \bibinfo{pages}{373--384}.
\newblock


\bibitem[\protect\citeauthoryear{Yen, Huang, Dai, Ravikumar, Dhillon, and
  Xing}{Yen et~al\mbox{.}}{2017}]%
        {yen2017ppdsparse}
\bibfield{author}{\bibinfo{person}{Ian~EH Yen}, \bibinfo{person}{Xiangru
  Huang}, \bibinfo{person}{Wei Dai}, \bibinfo{person}{Pradeep Ravikumar},
  \bibinfo{person}{Inderjit Dhillon}, {and} \bibinfo{person}{Eric Xing}.}
  \bibinfo{year}{2017}\natexlab{}.
\newblock \showarticletitle{Ppdsparse: A parallel primal-dual sparse method for
  extreme classification}. In \bibinfo{booktitle}{\emph{Proceedings of the 23rd
  ACM SIGKDD International Conference on Knowledge Discovery and Data Mining}}.
  \bibinfo{pages}{545--553}.
\newblock


\bibitem[\protect\citeauthoryear{Yen, Huang, Ravikumar, Zhong, and Dhillon}{Yen
  et~al\mbox{.}}{2016}]%
        {yen2016pd}
\bibfield{author}{\bibinfo{person}{Ian En-Hsu Yen}, \bibinfo{person}{Xiangru
  Huang}, \bibinfo{person}{Pradeep Ravikumar}, \bibinfo{person}{Kai Zhong},
  {and} \bibinfo{person}{Inderjit Dhillon}.} \bibinfo{year}{2016}\natexlab{}.
\newblock \showarticletitle{Pd-sparse: A primal and dual sparse approach to
  extreme multiclass and multilabel classification}. In
  \bibinfo{booktitle}{\emph{International Conference on Machine Learning}}.
  \bibinfo{pages}{3069--3077}.
\newblock


\bibitem[\protect\citeauthoryear{You, Zhang, Wang, Dai, Mamitsuka, and Zhu}{You
  et~al\mbox{.}}{2019}]%
        {you2019attentionxml}
\bibfield{author}{\bibinfo{person}{Ronghui You}, \bibinfo{person}{Zihan Zhang},
  \bibinfo{person}{Ziye Wang}, \bibinfo{person}{Suyang Dai},
  \bibinfo{person}{Hiroshi Mamitsuka}, {and} \bibinfo{person}{Shanfeng Zhu}.}
  \bibinfo{year}{2019}\natexlab{}.
\newblock \showarticletitle{Attentionxml: Label tree-based attention-aware deep
  model for high-performance extreme multi-label text classification}. In
  \bibinfo{booktitle}{\emph{Advances in Neural Information Processing
  Systems}}. \bibinfo{pages}{5820--5830}.
\newblock


\bibitem[\protect\citeauthoryear{Yu, Zhong, and Dhillon}{Yu
  et~al\mbox{.}}{2020}]%
        {pecos2020}
\bibfield{author}{\bibinfo{person}{Hsiang-Fu Yu}, \bibinfo{person}{Kai Zhong},
  {and} \bibinfo{person}{Inderjit~S Dhillon}.} \bibinfo{year}{2020}\natexlab{}.
\newblock \showarticletitle{PECOS: Prediction for Enormous and Correlated
  Output Spaces}.
\newblock \bibinfo{journal}{\emph{arXiv preprint arXiv:2010.05878}}
  (\bibinfo{year}{2020}).
\newblock


\end{thebibliography}

\newpage
\appendix

\section{Implementation Details}
\label{sec:implementation_details}

\subsection{Evaluation metric details}
Mean Reciprocal Rank (MRR) is the average of reciprocal ranks of the ground-truth next query in the ranked list of top-$k$ suggestions.
\[MRR = \frac{1}{N}\sum_{i=1}^{N} \frac{1}{r_i}\]
where $r_i$ is rank of ground-truth next query $(\nextQuery_{i})$ in top-$k$ suggestions for $i^\text{th}$ data point. If $\nextQuery_{i}$ is not present in top-$k$ suggestions, then $r_i = \infty$.

\RRBLEU is defined as reciprocal rank weighted average of the \BLEU score~\cite{papineni2002bleu} between the ground-truth next query and suggestions. 
\[ \RRBLEU = \frac{1}{N}\sum_{i=1}^{N} \frac{\sum_{j=1}^{k} \frac{1}{j} \BLEU(\nextQuery_{i}, \hat{q}_{i,j})}{\sum_{j=1}^{k} \frac{1}{j}} \]
where $\hat{q}_{i,j}$ is $j^\text{th}$ suggestion for $i^\text{th}$ data point, and $\nextQuery_{i}$ is the ground-truth next query for $i^\text{th}$ data point.
\BLEU score between $\nextQuery_{i}$ and $\hat{q}_{i,j}$ is computed using token-level n-gram overlap.
We use up to 4-gram and use smoothing method-1 from~\citet{chen2014systematic} to handle zero n-gram overlaps.

\subsection{Implementation details for all models}
Generative seq2seq-GRU model is trained to maximize the likelihood of ground-truth next 
query given the previous query. At test time, 
the top-$k$ next query suggestions are generated using beam search with 
the constraint that the generated suggestions start with the input 
prefix $\prefix$, and the maximum length of a generated sequence is 10 tokens. 
We use a beam width of 16 and use top-10 of generated queries as auto-complete suggestions.
We use SentencePiece tokenization~\cite{kudo2018sentencepiece} to 
create a shared vocabulary for encoder and decoder with a vocabulary 
size of 32,000. We use 256 dim input embedding, 1,000 dim hidden 
embedding, and share embeddings between encoder 
and decoder models. We optimize using Adam~\cite{kingma2014adam} with learning rate = 0.001.
 
For all variants of the proposed \our models, we encode the previous query ($\prevQuery$) using 
word token based unigram tfidf vectorization, and we encode prefix ($\prefix$) and label (next query $\nextQuery$)
using tfidf vectorization with character unigrams, bigrams and trigrams. The vectorizers are trained using scikit-learn's TFIDF module~\cite{pedregosa2011scikit}. The Position-Weighted vectorization scheme is implemented separately by modifying scikit-learn. 

For training \our models, we use PECOS, a modular Extreme Multi-Label Ranking (XMR) framework. 
We use our proposed label indexing variants encoded in tree data structures for the indexing step in PECOS. We use the linear 1-vs-all classifiers at each internal node for matching and linear 1-vs-all classifiers per label for ranking (XLINEAR mode in PECOS package). The individual classifiers are trained using $\ell_2$ hinge loss commonly used in linear SVM training with real positive examples and derived hard negative examples. For more details, we refer the reader to~\cite{pecos2020}. 

Code and scripts to reproduce results for \our models is available at \href{https://github.com/amzn/pecos}{https://github.com/amzn/pecos}.

\end{document}